\documentclass[aps,prl,twocolumn,showpacs,groupedaddress]{revtex4}

\usepackage{graphicx}

\begin{document}

\title{ Quasi-Particle Spectra, Charge-Density-Wave, Superconductivity and 
Electron-Phonon Coupling in 2H-NbSe$_2$}   
\author{T. Valla}
 \email{valla@bnl.gov}
\author{A. V. Fedorov}
 \altaffiliation[Permanent address: ]{Advanced Light Source, LBNL, Berkeley, CA 94720}
\author{P. D. Johnson}
\affiliation{ Department of Physics, Brookhaven National Laboratory, Upton, NY, 
11973-5000}
\author{P-A. Glans}
\author{C. McGuinness}
\author{K. E. Smith}
\affiliation{ Department of Physics, Boston University, 590 Commonwealth Avenue, 
Boston, Massachusetts 02215}
\author{E. Y. Andrei}
\affiliation{ Department of Physics, Rutgers University, 126 Frelinghuysen Road, Piscataway, NJ 08854}
\author{H. Berger}
\affiliation{ Institute of Physics of Complex Matter, EPFL, Lausanne, 
Switzerland}
\date{\today}

\begin{abstract}
High-resolution photoemission has been used to study the electronic structure of 
the charge density wave and superconducting dichalcogenide, 2H-
NbSe$_2$. From the extracted self-energies, important components of the 
quasiparticle interactions have been identified. In contrast to previously 
studied TaSe$_2$, the CDW transition does not affect the electronic properties 
significantly. The electron-phonon coupling is identified as a dominant 
contribution to the QP self-energy and is shown to be very anisotropic ($k$-
dependent) and much stronger than in TaSe$_2$.
\end{abstract}


\maketitle

The family of layered 2H dichalcogenides represents an interesting system in 
which charge-density wave (CDW) order co-exists with the superconductivity (SC) 
\cite{Wilson,Castro}. The fact that the CDW transition temperature decreases, while the 
superconducting critical temperature ($T_C$) increases from TaSe$_2$ through 
TaS$_2$ and NbSe$_2$ to NbS$_2$ suggests that these two order parameters 
represent competing ground states. Indeed, it has been found that in TaS$_2$ and 
NbSe$_2$, $T_C$ increases under pressure while $T_{CDW}$ decreases 
\cite{Smith,Moline}. After CDW order disappears, $T_C$ remains approximately 
constant. In NbS$_2$, the system without CDW order, $T_C$ is insensitive to 
pressure. Although various anomalies, including an apparent anisotropy of the 
superconducting gap, have been observed \cite{Garoche,Graebner,Corcoran,Yokoya}, 
it is generally believed that superconductivity in the dichalcogenides is of 
conventional BCS character, mediated by strong electron-phonon coupling 
\cite{Corcoran}. However, consensus on the exact mechanism that drives the 
system into the CDW state has still not been reached. Some authors 
\cite{Wilson,Doran,Wilson1,Castro} argue, in analogy with a Pierls transition in one-
dimensional systems, that the CDW transition is driven by a FS instability 
(nesting), where some portions of the FS are spanned by a CDW vector- $q_{CDW}$. 
In another scenario, the CDW instability is induced by the nesting of van Hove 
singularities (saddle points in the band structure with high density of states) 
if they are within a few $k_BT_{CDW}$ of the Fermi energy \cite{Rice}. 
The Fermi surface of the 2H dichalcogenides is rather complicated, being 
dominated by several open (2D-like) sheets and with one small 3D S(Se)-derived pancake-like FS \cite{Corcoran}. In 
such a situation, one may anticipate anisotropic properties and in particular, 
an anisotropic electron-phonon coupling. The resistivity anisotropy, of the 
order of 10-50, is much smaller than in layered oxides, indicating a substantial inter-layer hopping \cite{Laszlo}. 
Transport properties show relatively small anomalies at $T_{CDW}$, suggesting 
that only a small portion of the FS becomes gapped in the CDW state. In 
addition, the 2H dichalcogenides become better conducting in the CDW state, 
indicating a higher degree of coherence.

We have previously studied the quasi-particle (QP) self-energy for the first member 
of the series, TaSe$_2$, and found that it was strongly influenced by the CDW 
transition \cite{Valla-TaSe}. Sharp structure in both the dispersion ("kink") 
and the scattering rate at $\sim$70 meV in the CDW state weakened and shifted to 
$\sim$35 meV upon transition into the normal state. Similar "kinks", have been previously 
identified in conventional metals and attributed to electron-phonon coupling \cite{Mo-Be}. 
As the energy scale of the "kink" in the CDW state of TaSe$_2$
was too large for phonons, we suggested that the excitation was electronic in 
origin: a fluctuation of the CDW order parameter. In this letter, we show that 
the self-energy of NbSe$_2$ ($T_{CDW}\approx35$ K and $T_C=7.2$ K) is less 
sensitive to the CDW transition and that it is dominated by electron-phonon 
coupling. Measured directly for the first time, the coupling is very 
anisotropic, with the largest value on the inner K-H centered sheet. 

\begin{figure*}
\includegraphics{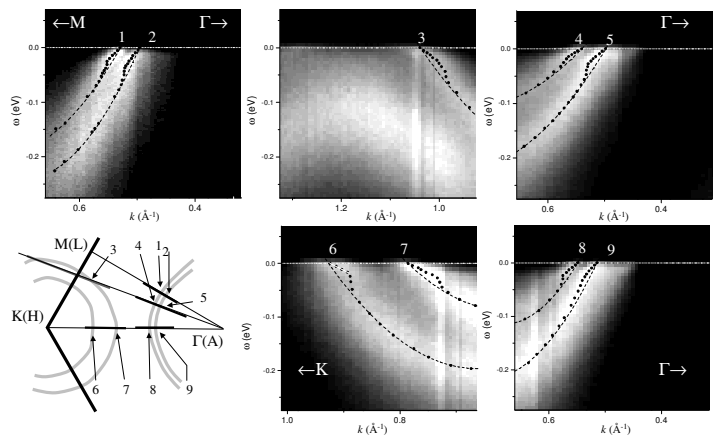}
\caption{\label{Fig:1}
The photoemission intensities in the CDW state at T=10K for several momentum lines indicated in the schematic view of the Brillouin zone (lower left panel) by the dark-gray lines. The light-gray lines represent Nb-derived Fermi sheets. The nine Fermi points are numbered.
The MDC derived dispersions are represented by full circles. The high-energy part of the
dispersions is fitted with a second-order polynomial (dashed lines) and the low energy part is fitted 
with straight lines.}
\end{figure*}

The experiments reported here were carried out on a high-resolution 
photoemission facility based on undulator beam line U13UB at the National 
Synchrotron Light Source and a Scienta SES-200 electron 
spectrometer, which simultaneously collects a large angular window $\approx14^\circ$ 
of the emitted photoelectrons. The combined 
instrumental energy resolution was set to 
$\sim$4 meV for low temperature studies ($T<15$K) and to $\sim$6 meV elsewhere. 
The angular resolution was better than $\pm 0.1^\circ$ translating into a 
momentum resolution of $\pm 0.0025 \AA^{-1}$ at the 15.2 eV photon energy used 
in the present study. Samples, grown by the standard iodine vapor transport method, were mounted on a liquid He cryostat and cleaved 
\textit{in-situ} in the UHV chamber with base pressure $3\times 10^{-9}$ Pa. The 
temperature was measured using a calibrated silicon sensor mounted near the 
sample. 

Figure \ref{Fig:1} shows the photoemission intensity, recorded at $T=$10 K, as a 
function of binding energy and momentum along three different momentum lines in 
the Brillouin zone. Nine Fermi crossings 
are included: three pairs on the double-layer split Fermi sheets centered around 
$\Gamma$ and three crossings on the split sheets centered at the K point. A 
characteristic change in the QP velocity ("kink") can be easily identified in 
all crossings. The kinks are also accompanied by a sharp change in the QP widths 
at the "kink" energy. These observations are indicative of (bosonic) excitations 
interacting with the QPs. The "kink" occurs roughly at the energy of the 
excitation involved in the coupling. Compared to TaSe$_2$, the excitation 
spectrum is limited to significantly lower energies. It also appears that the 
kink is not unique; its strength and energy depend on \textbf{k}, being different 
for different crossings. To quantify this, we have used momentum distribution 
curves (MDC) to extract the band dispersions for all the states shown. The 
dispersion curves provide direct information on the real part of the self-
energy, Re$\Sigma$. As shown in the figure, the high-energy part of the 
extracted dispersion can be fitted with a parabola that crosses through $k_F$, 
whereas the low energy part ($\omega<15-20$ meV) is fitted with a straight line. 
If we assume that the parabola represents the "non-interacting" dispersion (i.e. 
the dispersion in the absence of interactions that cause the kink), then the 
slopes of these two lines at $\omega=0$ may be used to directly extract the 
coupling constant, $\lambda=v_F^0/v_F-1$, where $v_F^0$ is the "non-interacting" 
(bare) Fermi velocity and $v_F$ is the renormalized one. 

\begin{figure}
\includegraphics{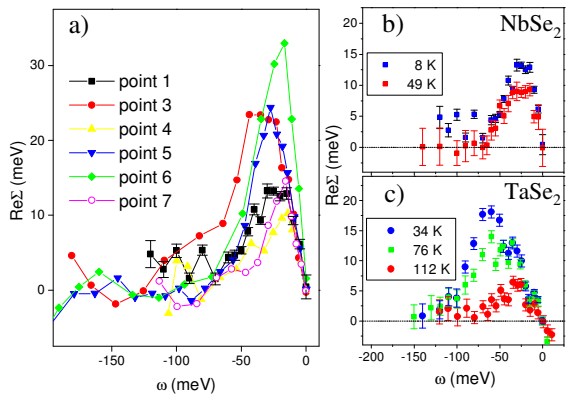}
\caption{\label{Fig:2}
(a) Real parts of self-energies Re$\Sigma$ extracted from measured dispersions from 
Fig. \ref{Fig:1} for several Fermi points. (b) Temperature dependence of Re$\Sigma$ for NbSe$_2$ 
for point 1 from Fig. \ref{Fig:1} and (c) for TaSe$_2$ (taken from ref. \cite{Valla-TaSe}) near the 
same region on the $\Gamma$-centered FS.}
\end{figure}

The "non-interacting" parabolas are further subtracted from the measured 
dispersions to extract Re$\Sigma(\omega)$. The results are shown in Fig. \ref{Fig:2}(a)
 for several crossings from Fig. \ref{Fig:1}. Re$\Sigma(\omega)$ gives the 
same coupling constant ($\lambda=-
[\partial(\text{Re}\Sigma)/\partial\omega]_0$), but also provides additional 
information about the spectrum of excitations interacting with the QPs. It is 
obvious from Fig. \ref{Fig:2}(a) that not only is the magnitude of Re$\Sigma$ 
different for different states, but also the peaks are at different energies, 
ranging from $\sim$13 meV to $\sim$35 meV. Various experimental and theoretical 
studies have shown that the phonon spectrum is fully consistent with these energies, with acoustical phonon branches laying below 
$\omega\sim$12 meV, and optical branches spanning the region 
$15<\omega<40$ meV \cite{phonons}. Shifts of the Re$\Sigma$ maxima would further suggest that some electronic 
states are coupled predominantly to acoustic modes while others couple more 
strongly to the optical modes, even though the states are sometimes very close 
in momentum (compare points 4 and 5, for example). A strong $k$-dependence of $\Sigma$ would 
complicate the MDC line-shape in the energy region where the $k$-dependence 
exists. Although we have detected some deviation from Lorentzian lineshapes, we 
were not able to precisely determine the MDC's line-shape due to the overlap of bi-layer split states. 
It is interesting that 
in spite of these differences in $\Sigma$, the resulting coupling constant does 
not vary much, $\lambda\sim0.85\pm0.15$, within the experimental uncertainty. 
The only exception is the inner K-centered sheet (point 6), where $\lambda\sim1.9\pm0.2$. 
We have completed several measurements on different samples, always with similar 
results for $\lambda$ at, or close to that Fermi point. This seemingly too 
large coupling constant is actually in a good agreement with the large measured 
value of linear specific heat coefficient, $\gamma\approx18.5$ mJmol$^{-1}$K$^{-
2}$ \cite{Garoche,specific-heat}, which is proportional to the renormalized density of 
states (DOS) at the Fermi level, $N(0)(1+\lambda)$, through 
$\gamma=(1/3)\pi^2k_B^2N(0)(1+\lambda)$. Band structure calculations give the 
"bare" DOS $N(0)\sim2.8$ states eV$^{-1}$ unit cell$^{-1}$ \cite{Corcoran}, suggesting 
$\lambda\sim1.8$. However even this might be an underestimate for our state as 
$\gamma$ measures an average over the FS, weighted by each state's DOS. A 
similar value for $\lambda$ is obtained from $c$-axis optical conductivity \cite{Sasa} 
suggesting that the $c$-axis transport is probably dominated by the K-H centered 
cylinders with largest warping.

It is instructive that in TaSe$_2$ the CDW gap opens up in the same region of 
the FS \cite{ARPES-Kpoint}, while the $\Gamma$-A centered Fermi cylinders remain 
ungapped, and gain coherence in the CDW state \cite{Valla-TaSe}. Therefore, it seems plausible 
that both the superconductivity and the CDW state originate from the inner K 
sheet and are driven by strong electron-phonon coupling. This seems to be in line 
with the original suggestion of Wilson \cite{Wilson1} that the self-nesting of the 
inner K sheet drives the CDW in 2H dichalcogenides. A lack of CDW gap on the $\Gamma$ centered 
sheets in all of 2H dichalcogenides studied in ARPES suggests 
that these sheets support neither the self-nesting nor the nesting which would mix them with the
K-centered sheets. In particular, the f-wave symmetry of the CDW gap \cite{Castro} may be ruled out. 
The relative strength of the CDW and superconducting ordering is determined 
by the nesting properties of the inner K cylinder, while the upper limit for $T_C$ (when the CDW is destroyed by applying 
pressure, for example) is given by $\lambda$. Nesting weakens with increasing 
3D character (increased warping with $k_Z$) under pressure and on moving from 
TaSe$_2$ to NbS$_2$. $\lambda$ increases from TaSe$_2$ to NbSe$_2$ \cite{NbS2} 
and is essentially pressure independent.
In agreement with previous ARPES studies \cite{Straub,Yokoya,Tonjes}, we do not see a CDW gap in NbSe$_2$, suggesting
that the nested portion of the FS is very small and was not sampled in any study. As there is a non-trivial $k_Z$-dispersion 
(warping), it is possible that the in-plane $k_F$ might be tuned 
into the nesting and that the gap opens only near certain $k_Z$. Note that the 
energy splitting between the double walled sheets is larger 
for K-centered sheets. A similar $k$-dependence is also expected for the interlayer hopping, $t_{\perp}$,
 that produces the warping. 
Additionally, as Fermi velocities are larger for $\Gamma$-
centered sheets, it is reasonable to expect that the in-plane $k_F$ varies with $k_Z$ much 
less on the $\Gamma$-cylinders than on the K-cylinders (the change in the in-plane 
Fermi momentum is approximately given by $\Delta k_F\propto t_{\perp}/v_F$). The 
measured FSs centered at $\Gamma$ are too large at the sampled $k_Z$, and we do not 
expect them to ever reach the self-nesting condition $2k_F=q_{CDW}$. On the other 
hand, the inner K-centered sheet seems to be very close to producing the 
required nesting. A more detailed mapping is needed to explore the nesting 
properties and eventual CDW gap opening in this region. According to STM studies 
\cite{STM}, the CDW gap is large ($\Delta_{CDW}\sim35$ meV) and should be easily measurable 
in ARPES. 
The overall electronic properties in NbSe$_2$ are much less sensitive to the CDW transition 
than in TaSe$_2$. Even the CDW induced structure in the self-energy that existed in TaSe$_2$ is absent in NbSe$_2$. 
Both the "kink" and the scattering rate are remarkably insensitive to the CDW 
(See Fig. \ref{Fig:2}b). 

\begin{figure}
\includegraphics{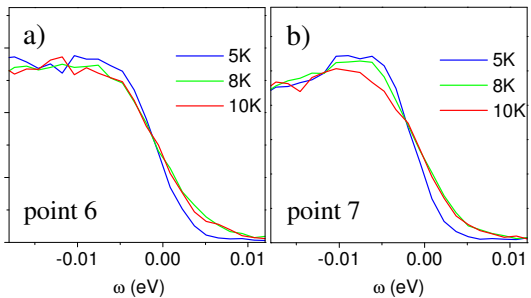}
\caption{\label{Fig:3}
Temperature dependence of the EDCs taken at Fermi points of the inner (a) 
and the outer (b) Fermi sheets centered at K. The superconducting gap, measured as the shift of the inflection point of the leading edge, is the same even though 
the measured $\lambda$ differs by a factor of 2.}
\end{figure}

An interesting question is whether the anisotropic electron-phonon coupling constant $\lambda$ would be projected into 
the magnitude of the superconducting gap. A recent photoemission study \cite{Yokoya} 
has shown that the superconducting gap is indeed anisotropic, being quite 
uniform, $\Delta\sim1$ meV, on the Nb-derived Fermi cylinders, but reduced 
beyond the experimental sensitivity ($\Delta\approx0$) on the 3D pancake-like 
FS. No data for the inner K-H cylinder has been reported. In Fig. \ref{Fig:3} we show the 
spectral intensity at two Fermi points, \textbf{6} and \textbf{7} for several temperatures. 
The coupling constant differs by a factor of $\sim2$ at these two points and yet, the superconducting gap,
 that clearly opens up below $T_C$ at both Fermi points, is the same within the experimental error. This is in contrast 
to MgB$_2$, where the "hot" regions are gapped by correspondingly larger gaps \cite{Tu}. 
The equalizing of the superconducting gap for states with different coupling strengths but similar symmetries represents a 
$k$-space analogous of the proximity effect \cite{proximity}.

NbSe$_2$ has several properties in common with other layered materials. One 
of the most obvious peculiarities is the fact that the transport becomes more 
metallic in the CDW state even though a portion of the FS is "destroyed" 
(gapped). We suggest that this is a manifestation of anisotropic Fermi surfaces where 
strongly coupled portions, or "hot spots", generally play a negative role in 
normal state conductivities, acting as scattering "sinks" for the remaining, less renormalized 
regions. These "hot spots" drive the system into an ordered state, but it is the 
"cold spots" that usually dominate the 
conductivities. This duality has recently been detected in MgB$_2$ where the 
normal state transport shows an extremely weak coupling, while the 
superconducting properties are dictated by strongly coupled portions on the FS 
\cite{Tu}. A Similar situation also exists in the cuprates where the in-plane normal 
state properties are dominated by near-nodal region \cite{Valla-SciencePRL}. The antinodal regions 
reduce the in-plane conductivities, while the opening of the pseudogap improves 
them. 

Another similarity with the cuprates exists in that the pseudogap ($T^*$) and 
$T_C$ lines in the $T-x$ phase diagram of cuprates show similar behavior to the 
$T_{CDW}$ and $T_C$ lines in the $T-pressure$ phase diagram of dichalcogenides. 
As $T^*(T_{CDW})$ goes down, $T_C$ increases with doping (pressure). This 
analogy suggests that the cuprate phase diagram might be shaped by similar competing orders. 
However, the analogy is no longer valid in the overdoped regime where $T_C$ turns back down, 
even when the pseudogap no longer exists, suggesting that some other, more exotic coupling mechanism 
might be acting in the cuprates. 
Here, we would like to point out that the high 
energy of the "kinks" observed in TaSe$_2$ \cite{Valla-TaSe} and recently in NbSe$_3$ \cite{NbSe3} rules 
out the phonon scenario and represents an unambiguous evidence that the "kinks" may indeed be caused by
 other mechanisms. Consequently, the "kinks" observed in the cuprates \cite{PDJ-Lanzara} do not 
necessarily reflect the electron-phonon coupling .

In conclusion, we have detected a strong anisotropy of the self-energy in a 
layered dichalcogenide, 2H-NbSe$_2$, with the electron-phonon coupling constant 
$\lambda$ ranging from 0.8 to 1.9 on Nb-derived sheets. The strongest coupling 
has been found on the inner K-H cylinder, which plays a central role in both CDW 
and SC transitions in 2H-dicalcogenides. The anisotropy in coupling strength 
does not induce the anisotropy in the SC gap. 
\begin{acknowledgments}
The authors would like to acknowledge useful discussions with Tim Kidd, Sa\v sa 
Dordevi\'c, Laszlo Forr\'o and Antonio Castro Neto. The BNL program was supported by the US DOE under contract number DE-AC02-98CH10886. The Boston University program was supported by the US DOE under DE-FG02-98ER45680. 
The sample preparation in Lausanne was supported by the NCCR research
pool MaNEP of the Swiss NSF.
\end{acknowledgments}


\begin{references}
\bibitem{NbS2} 2H-NbS$_2$ has slightly lower $T_C$ ($\sim6.3$ K), suggesting a smaller coupling constant.
\end{references}
\end{document}